\documentclass[pra,aps,amssymb,twocolumn,showpacs]{revtex4}
\usepackage{color}
\usepackage{epsf}
\newcounter{abcd}
\newcommand{\be}{\begin{equation}}
\newcommand{\ee}{\end{equation}}
\newcommand{\bd}{\begin{displaystyle}}
\newcommand{\ed}{\end{displaystyle}}
\begin{document}
\oddsidemargin=0cm
\topmargin=0cm
\title{Synchronization of internal and external degrees of freedom of atoms in
a standing laser wave}
\author{V.Yu. Argonov and S.V. Prants}
\affiliation{Laboratory of Nonlinear Dynamical Systems,\\
V.I.Il'ichev Pacific Oceanological Institute of the Russian Academy of
Sciences,\\
690041 Vladivostok, Russia}
\date{\today}
\begin{abstract}
We consider dissipative dynamics of atoms in a strong standing laser
wave and find a nonlinear dynamical effect of synchronization between
center-of-mass motion and internal Rabi oscillations. The synchronization
manifests itself in the phase space as limit cycles which may have
different periods and riddled basins of attraction. The effect can be
detected in the fluorescence spectra of atoms as equidistant sideband
frequencies with the space between adjacent peaks to be inversely
proportional to the value of the period of the respective limit cycle.
With increasing the intensity of the laser field, we observe numerically
cascades of bifurcations that eventually end up in settling  a
strange chaotic attractor. A broadband noise is shown to destroy a fine
structure of the bifurcation scenario, but prominent features of
period-1 and period-3 limit cycles survive under a weak noise.
The character of the atomic motion is analyzed with the help of the friction
force whose zeroes are attractor or repellor points in the velocity space.
We find ranges of the laser parameters where the atomic motion resembles
a random  but deterministic walking of atoms erratically jumping between
different wells of the optical potential. Such a random walking is shown to
be fractal in the sense that the measured characteristic of the motion,
time of exit of atoms from a given space of the standing wave, is
a complicated function that has a self-similar structure with singularities
on a Cantor set of values of one of the control parameters.
\end{abstract}
\maketitle

\section{Introduction}

Light exerts mechanical forces on neutral atoms whose origin can be explained
as in the particle as in the wave pictures of the particle-wave duality. In the
first one, light is a collection of photons carrying energy, angular momentum,
and momentum. When interacting with light, atoms change not only their internal
electronic states but also their external translational states, the process
known as a photon recoil. If one considers light as a wave, the mechanical
force is produced by the interaction between inhomogeneous electromagnetic
wave and the atomic dipole moment.

In this paper we consider atoms confined in a cavity or in a standing wave made
up of two counter-propagating laser waves. There is a vast literature on the
various aspects of the topic of mechanical action of light on atoms (see, for
example, some reviews and books \cite{ML, KSY, CDG, S, ASM}). In the simplest
case of a two-level atom interacting with a single mode of an one-dimensional
standing wave in a cavity, there are, at least, three strongly coupled
subsystems: internal and external atomic degrees of freedom and field degrees
of freedom. Recently, it has been shown theoretically and numerically that
Hamiltonian dynamics of a single atom in a standing light wave demonstrates
a variety of new dynamical effects such as correlations between Rabi oscillations
and atomic translational motion \cite{V98, AP03}, the Doppler-Rabi resonance
\cite{AP03, UKP03}, Hamiltonian chaos \cite{PK01, PS01}, L\'evy flights
\cite{PEZ02, P02, AP03}, and atomic dynamical fractals \cite{P02, AP03, PU03}.
The semiclassical and quantum theory of these effects has been developed in the
cited papers.

In the present paper we study dissipative dynamics of two-level atoms in a
standing wave pumped by a laser and report on a nonlinear dynamical effect
of synchronization between internal and external atomic degrees of freedom.
Light-induced forces cause an atom either to be trapped in a well of the
optical potential or to fly through the standing wave. The mechanical
atomic oscillator forces the Rabi one to oscillate with the same rhythm.
Synchronization occurs after a transition time and manifests itself in the
phase space as limit cycles which may have different periods depending on
the detuning $\delta$ between the atomic transition, $\omega_a$, and
standing-wave frequencies, $\omega_f$, and on the average number of photons in
the wave $n$. Synchronization and the period of the respective limit cycle
can be detected by measuring fluorescence spectra of atoms. Synchronized ballistic
motion with a period-m limit cycle should generate sidebands with the frequencies,
$\omega_f\pm jk_f v_s/m\ (j=1,2,...)$ in the respective spectra (where
$ v_s$ is a quasistationary atomic velocity), whereas trapped atoms, oscillating
under a period-m limit cycle, generate sidebands $\omega_f\pm j\Omega_\xi
/m$,
where $\Omega_\xi$ is the respective frequency of nonlinear synchronized oscillations.
Asynchronous chaotic motion generates a broaden spectrum.

This work is organized as follows: in \setcounter{abcd}{2}Sec.~\Roman{abcd}
the model of an atom interacting
with a standing wave in a driven cavity is introduced, and semiclassical
equations of motion including spontaneous relaxation, cavity damping, and
pumping are derived. In the strong-field limit and under a resonance of
a laser
with the standing wave, the equations are reduced to a five-dimensional
dissipative nonlinear system with coupled mechanical and Bloch variables. In
\setcounter{abcd}{3}Sec.~\Roman{abcd} we discuss briefly two simple
regimes of atomic motion: shallow
oscillations of an atom trapped in the optical potential and ballistic
motion under
an influence of the light-induced friction force. The friction force is
analytically calculated for thermal atoms with a large detuning $\delta$. In
\setcounter{abcd}{4}Sec.~\Roman{abcd} we report on synchronization of
the atomic Rabi oscillations with
the momentum oscillations, give respective approximated analytic solutions
for period-1 limit cycles, show that basins of attraction of different
limit cycles are riddled, and compute spectra of atomic fluorescence.
In \setcounter{abcd}{5}Sec.~\Roman{abcd} we demonstrate a rich scenario
of bifurcations in dependence on the
average number of photons in the standing wave. The scenario contains
coexisting limit cycles of different periods, period doubling, and cascades
of different bifurcations on the road to a strange chaotic attractor.
\setcounter{abcd}{6}Sec.~\Roman{abcd} contains
numerical results demonstrating atomic dynamical fractals in the regime of
random walking of an atom in a standing wave and a comparison between
dissipative and Hamiltonian fractals.

\section{Equations of motion}

Let us consider a two-level atom with the electronic transition frequency
$\omega_a$ and mass $m_a$ in a 1D-cavity which sustains a single cosine-like
standing-wave mode along the axis $r$ with the wave number $k_f$ and the
frequency $\omega_f$. The cavity mode is supposed to be pumped by a classical
laser field with the amplitude $\varepsilon$, phase $\varphi$, and frequency
$\omega_l$. The total Hamiltonian
\be
\label{1}
\hat H=\hat H_0+\hat H_l
\ee
consists of the standard cavity QED Hamiltonian
\begin{eqnarray}
\label{2}
\hat H_0=\frac{\hat p^2}{2m_a}+\frac{1}{2}\hbar\omega_a
\hat\sigma_z+\hbar\omega_f\left(\hat a^\dagger\hat a+\frac{1}{2}\right)
-\nonumber\\
-\hbar\Omega_0\cos(k_f\hat r)
(\hat a^\dagger\hat\sigma_-+\hat a\hat\sigma_+)
\end{eqnarray}
and the laser-mode interaction Hamiltonian
\begin{equation}
\label{3}
\hat H_l=\hbar\varepsilon\hat a\exp i(\omega_l t-\varphi)+\mbox{H. c.}
\end{equation}
The atom may emit and absorb photons from the cavity mode. Both the processes
change internal atomic and field states. The external or translational atomic
states are changed as well because of a photon recoil. The laser field supplies
an energy to overcome atomic and mode dissipation. All the processes are
presented in the Hamiltonians. The operators $\hat r$ and $\hat p$ with the
commutation rule $[\hat r, \hat p]=i\hbar$ describe atomic position and momentum,
respectively. The Pauli atomic operators
$\hat\sigma_\pm\equiv\frac{1}{2}(\hat\sigma_x\pm i\hat\sigma_y)$ and $\hat\sigma_z$
with the commutation rules $[\hat\sigma_\pm,\hat\sigma_z]=\mp 2\hat\sigma_z$,
$[\hat\sigma_+,\hat\sigma_-]=\hat\sigma_z$ describe electronic atomic transitions.
The boson operators $\hat a$ and $\hat a^\dagger$ with the commutator
$[\hat a, \hat a^\dagger]=1$ describe the cavity mode. The atom-field coupling
is characterized by an amplitude of the single-photon Rabi frequency $\Omega_0$.

It is convenient to choose the following combinations of operators:
$\hat e\equiv\hat a^\dagger \exp (-i\omega_f t)+\hat a \exp (i\omega_f t)$,
$\hat g\equiv i[\hat a^\dagger \exp (-i\omega_f t)-\hat a \exp (i\omega_f t)]$,
$\hat x \equiv \hat \sigma_- \exp (i\omega_f t) + \hat \sigma_+ \exp
(-i\omega_f t)$, $\hat  y \equiv i[\hat \sigma_- \exp (i\omega_f t) -
\hat \sigma_+ \exp(-i\omega_f t)]$,
$\hat\sigma_z$, $\hat r$, and $\hat p$, for which the respective Heisenberg
equations can be derived using the Hamiltonians (\ref{2}) and (\ref{3}) in
the frame rotating with the frequency $\omega_f$

\begin{equation}
\begin{array}{l}
\bd\frac{d\hat r}{dt}=\frac{\hat p}{m}\ed, \nonumber \\
\\
\bd\frac{d\hat p}{dt}=\frac{\hbar k_f\Omega_0}{2}(\hat g\hat y-\hat e\hat x)
\sin(k_f\hat r),\ed\nonumber\\
\\
\bd\frac{d\hat e}{dt}=\Omega_0\hat y\cos(k_f\hat r)+2\varepsilon
\sin(\omega_ft-\omega_lt+\varphi)-2\Gamma_f\hat e,\ed\nonumber\\
\\
\bd\frac{d\hat g}{dt}=\Omega_0\hat x\cos(k_f\hat r)-2\varepsilon
\cos(\omega_ft-\omega_l t+\varphi)-2\Gamma_f\hat g,\ed\label{5}\\
\\
\bd\frac{d\hat x}{dt}=(\omega_f-\omega_a)\hat y+\Omega_0\hat\sigma_z
\hat g\cos(k_f\hat r)-\frac{1}{2}\Gamma_a\hat x,\ed\nonumber\\
\\
\bd\frac{d\hat y}{dt}=(\omega_a-\omega_f)\hat x+\Omega_0\hat\sigma_z
\hat e\cos(k_f\hat r)-\frac{1}{2}\Gamma_a\hat y ,\ed\nonumber\\
\\
\bd\frac{d\hat\sigma_z}{dt}=-\Omega_0(\hat g \hat x+\hat e\hat
y)\cos(k_f\hat r)-\Gamma_a(\hat \sigma_z+1),\ed\nonumber
\end{array}
\end{equation}
where $\Gamma_f$ and $\Gamma_a$ are the rate of dissipation of cavity photons
and spontaneous emission rate, respectively, to be added to the respective
equations phenomenologically.
After averaging over an initial quantum state, which is supposed to be a
product of the field, internal, and external atomic states, we get the respective
c-number equations of motion. Assuming a large number of photons in the cavity
mode, $n=(e^2+g^2)/4\gg 1$, and large values of the atomic momentum as compared with the
photon momentum $\hbar k_f$, we adopt the semiclassical approximation. It
results in the following normalized closed set of ODEs:
\begin{equation}
\begin{array}{lll}
\dot \xi =\alpha p,\\
\dot p = (1/2)(gy-ex)\sin\xi,\\
\dot e = y\cos\xi-2\gamma_f e+2E\sin(\Delta\tau+\phi),\\
\dot g = x\cos\xi-2\gamma_f g-2E\cos(\Delta\tau+\phi),\label{6}\\
\dot x=\delta y+zg\cos\xi-(1/2)\gamma_a x,\\
\dot y=-\delta x+ze\cos\xi-(1/2)\gamma_a y,\\
\dot z= -(gx+ey)\cos\xi-\gamma_a(z+1),
\end{array}
\end{equation}
where dot denotes differentiation with respect to the dimensionless time
$\tau\equiv\Omega_0t$ and $E\equiv\varepsilon/\Omega_0$. Here
$\xi\equiv k_f<\hat r>$, $z\equiv\ <\hat\sigma_z>$, $p=\ <\hat p>/\hbar k_f$,
and the other variables are the expectation values of the respective operators
written above. A few dimensionless frequencies define the atom-field dynamics:
$\alpha\equiv \hbar k^2_f/m_a\Omega_0$ is the photon recoil frequency,
$\delta\equiv(\omega_f-\omega_a)/\Omega_0$ the detuning between the frequencies
of the cavity mode and atomic electronic transition,
$\Delta\equiv(\omega_f-\omega_l)/\Omega_0$ the cavity-laser detuning, and
$\gamma_{a, f}$ the relaxation rates.

In fact, Eqs. (\ref{6}) are the coupled Hamilton-Maxwell-Bloch equations which
take into account not only light-induced mechanical forces acting on a
neutral atom but the back reaction of the atom on the cavity mode as well. In
a strong field the cavity mode reaches rapidly a regime with a large saturation
number of photons $n=(E/2\gamma_f)^2$. Throughout the paper we will consider
the case of exact resonance between the cavity mode and the laser, $\Delta=0$,
and a specified phase $\varphi=\pi/4$. It follows the reduced set of our basic
equations
\begin{equation}
\begin{array}{lll}
\dot \xi =\alpha p,\\
\dot p = -u\sin\xi,\\
\dot u =\delta v-(1/2)\gamma_a u,\label{7}\\
\dot v=-\delta u+2nz\cos\xi-(1/2)\gamma_a v,\\
\dot z= -2v\cos\xi-\gamma_a(z+1),
\end{array}
\end{equation}
where we introduce new variables $u\equiv(ex-gy)/2=\sqrt{n/2}\ (x+y)$
and $v\equiv(gx+ey)/2=\sqrt{n/2}\ (y-x)$ which are proportional to the
quadrature components of the atomic electric dipole moment. Namely, the
system (\ref{7}) will be treated throughout the paper. Eqs. (\ref{7}) are
written in the form convenient to the comparison with equations (\ref{3})
in Ref. \cite{AP03} which describe Hamiltonian self-consistent atom-field
dynamics in a high-Q cavity without dissipation and a pumping field. In fact,
we need no cavity in the strong field regime for which Eqs. (\ref{7}) are valid.
A standing wave may be created by two counter propagating laser beams in
a free space. Real laser beams produce, of course, a three dimensional standing
wave, and we simplify the geometry to simplify the analysis.

\section{Center-of-mass motion in a standing wave}
\subsection{Shallow oscillations in a well of the optical potential:
steady-state solutions}

An atom moves in an optical potential $\Pi=\int u\sin\xi d\xi$, whose form can
be easily found in the steady state approximation when the atom is assumed to
move so slowly that the Bloch variables $u$, $v$, and $z$ have a time to take
steady-state values. It is the case of rather cold trapped atoms oscillating in wells
of the optical potential that is valid if the normalized Doppler shift is much
smaller than the detuning and the relaxation rates, $|\alpha p|\ll|\delta|,
\gamma_{a, f}$.
With the time derivatives of the Bloch variables in Eqs. (\ref{7}) equal to
zero, it is easily to find their stationary values
\begin{equation}
\begin{array}{l}
\bd u_s=-\frac{2n\delta\cos\xi}{\delta^2+2n^2\cos^2\xi+\gamma_a^2/4}\ed, \nonumber \\
\\
\bd v_s=-\frac{\gamma_a n\cos\xi}{\delta_2+2n^2\cos^2\xi+\gamma_a^2/4},
\ed\label{8}\\
\\
\bd z_s=-\frac{1}{1+2n^2\cos^2\xi/(\delta^2+\gamma_a^2/4)}.\ed\nonumber
\end{array}
\end{equation}
Substituting the solution for $u_s$ in the second equation in (\ref{7}), we obtain
the gradient or dipole force acting on the atom
\be
\bd \dot p=\frac{n\delta\sin 2\xi}{\delta^2+2n^2\cos^2\xi+\gamma_a^2/4}\ed
\label{9}
\ee
and the optical potential
\be
\bd\Pi=\frac{\delta}{n}\ln\left(\frac{\gamma^2_a}{4}+\delta^2+2n^2\cos^2\xi
\right).\label{10}
\ed
\ee
Well known effects follow from this solution (see, for example, \cite{KSY, S}).
At exact resonance, $\delta=0$, the optical potential is zero, and the atom
moves with a constant velocity, equal to its initial value $p=p_0$.
At positive values of the detuning, $\delta>0$, atoms tend to go to nodes of
standing wave ($\xi=\pi/2+\pi m$), whereas at negative detuning, $\delta<0$,
they tend to go to antinodes ($\xi=\pi m$), where $m=0,1,...$. We recall that
our standing wave is of the form $-\cos\xi$.

\subsection{Ballistic motion and the friction force}

When an atom moves with a thermal average momentum  $ p_s$ through the
standing
wave, the Doppler shift exceeds relaxation rates, and the approximation of
slowly varying field amplitude is no valid. In the limit
$|\alpha p_s|>\gamma_{a, f}$, we will seek solutions for the Bloch variables
in the form
\begin{equation}
\begin{array}{l}
u=A\cos\alpha p_s\tau+B\sin\alpha p_s\tau, \label{11}\\
v=C\cos\alpha p_s\tau+D\sin\alpha p_s\tau.
\end{array}
\end{equation}
Let the atom reaches after a transition time a quasistationary momentum
$ p_s$  (around which its instant momentum oscillates slightly)
and does not experience Rabi oscillations due to a large detuning
$|\delta|\gg 1$, i. e. $z\simeq-1$. Substituting the solutions (\ref{11})
into Eqs. (\ref{7}), it is possible to find after some algebra an expression
for the force acting on the atom $f\equiv \dot p=-u\sin\xi$. After time
averaging
it, we get a dissipative or friction force, $F\equiv-d|\bar p|/d\tau$, which
has the form
\be
\label{12}
\bd
F\simeq-\frac{2n\delta\gamma_a\alpha| p_s|}
{(\gamma_a\delta)^2+
[(\alpha| p_s|)^2 - \delta^2+\gamma^2_a/4]^2}.
\ed
\ee
This approximated solution gives dispersion-like dependencies on the Doppler
shift $\alpha| p_s|$ and the detuning $\delta$. At negative detunings, the
friction force decelerates atoms whereas at positive detunings, it accelerates
them.

\begin{figure}[ht]
\begin{center}
\epsfxsize=8cm
\epsfbox{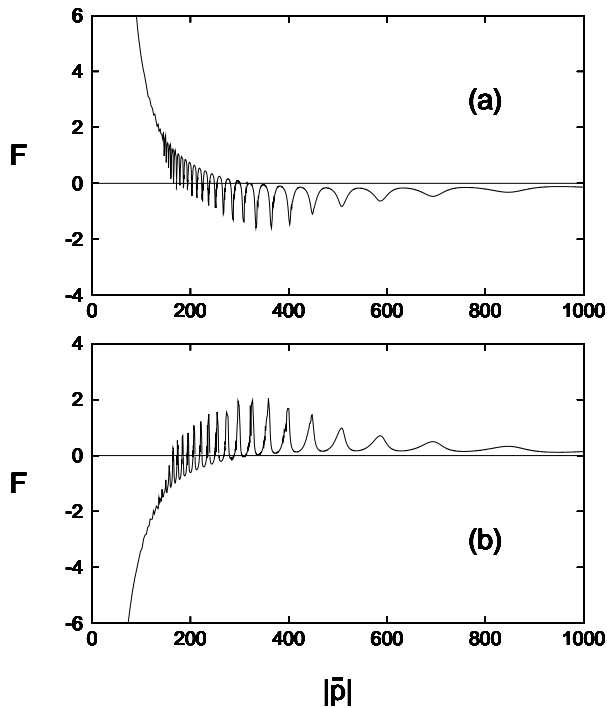}
\end{center}
\caption{\small Friction force $F$ in units of $\Omega_0\hbar k_f$ as a function
of the average atomic momentum $|\bar p|$ in units of the photon momentum
$\hbar k_f$. (a) Positive detuning $\delta=24$ and (b) negative detuning
$\delta=-24$.}
\label{fig01}
\end{figure}

In fact, the friction force $F$ depends on $\alpha| p_s|$ and $\delta$ in a more
complicated way. We integrate the basic Eqs. (\ref{7}) and plot in Fig.~1
its dependence on instant values of the atomic momentum $|\bar p|$,
averaged over a time of
flight between two successive nodes of the standing wave,
at $\delta>0$ (Fig.~1a) and $\delta<0$ (Fig.~1b). It is plotted for the
values of the momentum larger than a critical momentum $p_{cr}$ that is a minimal value of
$|\bar p|$ under which a ballistic motion is possible. What happens with atoms whose
momenta at some instant of time lie in a given range? At blue detunings,
$\delta>0$ , atoms with $|\bar p|<|p_a|$ are trapped, where $|p_a|$  is a
value of the average momentum which corresponds to the first zero of the
function $F(|\bar p|)$. The parts of this function between its successive
zeroes may be considered as averaged phase trajectories of atoms in the space
($|\bar p|, d|\bar p|/d\tau$). At positive values of the friction force,
atoms move with decreasing momentum, to the left on the plot. At $F <0$,
they move with increasing momentum, i.e. to the right on the plot. Thus,
zeroes of the function $F(|\bar p|)$ with positive
derivatives are attractors, whereas zeroes with negative
derivatives are repellors.

As it can be seen from Fig.~1b, at red detunings, $\delta <0$, repellors and
attractors
interchange their places as compared with Fig.~1a. It immediately means that
the friction at $\delta <0$ accelerates slow atoms and decelerates
fast ones which in a due time reach a quasistationary momentum $| p_s|$
and oscillate slightly in the velocity space around it (Fig.~2a). If the
initial momentum is sufficiently small, an atom may even change the direction
of motion, but a stabilization occurs in a due time. It is the
known effect of velocity grouping \cite{K74} that appears to be useful to
observe fluorescence spectra for an atomic ensemble with a velocity distribution
that is a more feasible task from the experimental point of view than observing
a single-atom fluorescence. Initially trapped atoms oscillate firstly in  wells
of the optical potential with increasing amplitudes, and then leave the wells
reaching after a transition time a grouping velocity (lower trajectory in Fig.~2a).

In difference from red detunings, where velocity grouping is global, at blue
detunings it is limited in the velocity space by a range of the atomic
momenta $|p_a|<|\bar p|<|p_b|$, containing zeroes of the function
$F(|\bar p|)$. The values of the momenta, $|p_a|$ and $|p_b|$,
corresponding to the first and last zeroes, are proportional to the
\begin{figure}[htb]
\begin{center}
\epsfxsize=8.5cm
\epsfbox{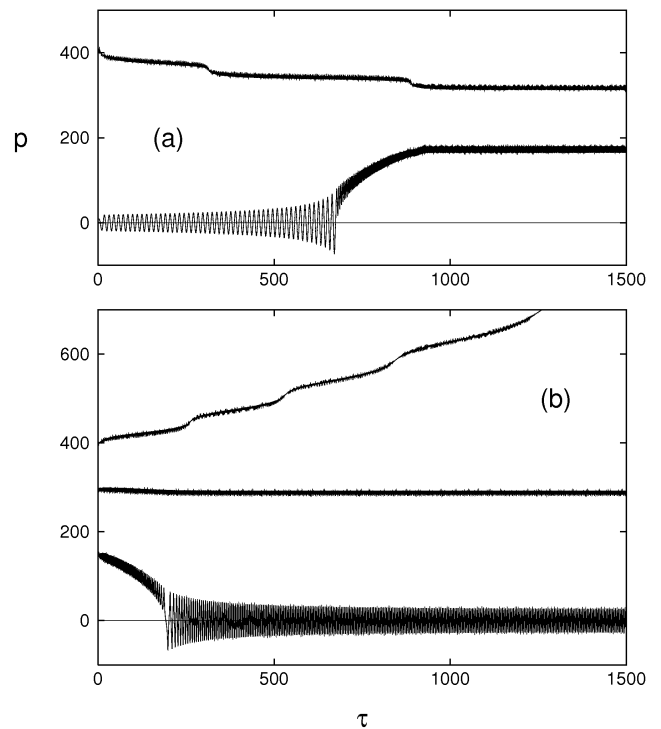}
\end{center}
\caption{\small Atomic momentum $p$ as a function of time $\tau$ in units of the
reciprocal of the single-photon Rabi frequency $\Omega_0$. (a) red detuning,
$\delta<0$: global velocity grouping for different values of the initial
momentum; (b) blue detunings, $\delta>0$: fast atoms are accelerated (upper
trajectory), slow atoms are decelerated (lower trajectory), and velocity
grouping occurs only for atoms in a range
$|p_a|\leqslant|\bar p|\leqslant|p_b|$.}
\label{fig02}
\end{figure}
detuning $\delta$ and depend on the other control parameters in an irregular way.
Thus, at $\delta >0$, slow atoms (with $|\bar p|\leq|p_a|\simeq 170$ in Fig.~1a) are
decelerated by the friction force and trapped in wells of the optical
potential eventually (see the lower trajectory in Fig.~2b). The friction
force for fast atoms ($|\bar p|\geq|p_b|\simeq 320$) is negative, and they are
accelerated by the field (the upper trajectory in Fig.~2b). The velocity
grouping at $\delta >0$ occurs only in the range
$170\lesssim|\bar p|\lesssim 320$.
A sample trajectory for such atoms is shown in Fig.~2b as a middle one.

\section{Synchronization}
\subsection{General features}

Our basic Eqs. (\ref{7}) describe two coupled oscillators, the external mechanical
one ($\xi, p$) and the internal Rabi oscillator ($u, v, z$) whose free frequencies
may differ in a few orders of magnitude in a strong field. The frequency of
small oscillations of an atom in a well of the optical potential can be
estimated to be $\alpha^{1/2}n^{1/4}$, whereas the free Rabi frequency,
$\sqrt{\delta^2+4n}$, is proportional to a square root of the average number
of photons $n$. The phenomenon we report in this section is synchronization of
internal and external degrees of freedom of an atom in a standing-wave field.
Synchronization is a fundamental nonlinear phenomenon to be observed for the first
time by Ch. Huygens in 1665 with his famous mechanical pendulum clock \cite{H}
and described theoretically by A. Andronov and A. Witt in 1930 \cite{AW} in
terms of limit cycles. In our case the effect consists in equality between
the frequency of oscillations of the atomic internal energy and the frequency
of variations of the atomic momentum as for trapped as for ballistic atoms.
The mechanical oscillator forces the Rabi one to oscillate with the same rhythm.
A standing wave with the spatial period $2\pi$ modulates with the Doppler
shift $\alpha p_s$ momentum of an atom, moving with an average momentum
$ p_s$.

\begin{figure}[htb]
\begin{center}
\epsfxsize=8.5cm
\epsfbox{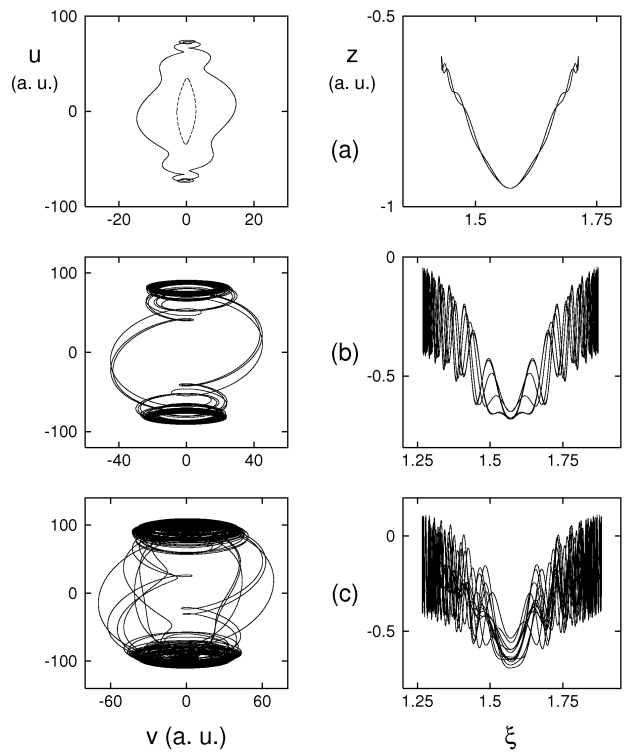}
\end{center}
\caption{\small Projections of the phase trajectories of a trapped atom onto two planes
($u, v$) and ($\xi, z$) at $\delta=24$: (a) period-1 limit cycle (dashed
line at $n=3000$ and solid line at $n=10000$), (b) period-3 limit cycle
($n=14846$), and (c) chaotic strange attractor ($n=24000$).
The atomic position $\xi$ is shown in units of $k_f^{-1}$.}
\label{fig03}
\end{figure}

Synchronization manifests itself in the phase space as limit cycles which may
have different periods. A limit cycle of period $m$ means a periodicity occurs
when the atom transverses $2m$ times the same node (being trapped in a well)
or different nodes (a ballistic atom) of the standing wave. We have found
limit cycles with m=$1,2,3,...,12$. In Fig.~3a and 3b we plot projections of
period-1 and period-3 limit cycles on the Bloch plane ($v, u$) and on the plane
($\xi, z$) of the internal and external variables at $\delta=24$, $\alpha=0.01$,
$\gamma_a=0.3$, and $p_0=60$. The size of a limit cycle is defined by the
amplitude of the respective synchronized oscillations, its form --- by the
spectrum of the oscillations, and the time of circulation of a phase point
over the cycle --- by the period of oscillations. The larger the parameter
of nonlinearity, $n$, the stronger the form of a limit cycle differs from the
harmonic one. In Fig.~3a the dashed line is a period-1 limit cycle with
$n=3000$ and the solid one is a cycle with $n=10000$.

Synchronization of internal and external atomic degrees of freedom is a common
feature of regular motion that occurs with different initial states and in
wide ranges of values of the control parameters. The only exception is the case
of exact resonance, $\delta=0$, when the optical potential vanishes and an atom
flies through the wave with a constant velocity, but its internal energy oscillates
with a modulated amplitude and frequency. Chaos destroys synchronization if one
understands under this name a frequency locking, where two frequencies become
rationally related. Synchronization of chaotic oscillations (if any) is beyond
the scope of this paper.

The time of settling limit cycles depends mainly on the action of the
force $f$ which accelerates or decelerates an atom during a transition
time $\tau_s$ before it reaches a quasistationary momentum $ p_s$. This
time can be estimated for atoms with velocities $p$ closed to $ p_s$
as follows. The force can be expanded in the Taylor series nearby the
point $ p_s$
\be
\Delta f =\Delta p f^\prime ( p_s) + ... ,\label{13}
\ee
where $\Delta p\equiv  p_s -p$, $\Delta f\equiv f( p_s) -f(p)$, and
$f^\prime\equiv df/dp$.  On the other hand, for small $\Delta p$,
$\Delta f \simeq  \Delta p/\tau_s$, and in the first order we get $\tau_s\simeq
[f^\prime ( p_s)]^{-1}$. It follows from (\ref{12}), the transition
time is a reciprocal of the average number of photons $n$.

\subsection{Approximated analytic solutions for period-1 cycles}

Depending on the values of the control parameters and the initial momentum,
atoms either are trapped in wells of the optical potential and oscillate
with $ p_s=0$, or they fly through the standing wave ($ p_s\ne 0$).
The frequency of synchronized motion is defined by the frequency of oscillations
of $p$ around $ p_s$. At $\delta>0$ and sufficiently large values of the
initial momentum $p_0$, atoms can move with an acceleration asymptotically
going to zero (Fig.~1a and Fig.~2b, the upper trajectory).

In order to find simple analytic solutions for period-1 limit cycle in the case
of ballistic motion, we put $\xi\approx\alpha p_s$ and
expand the variables $p$, $u$, $v$, and $z$ in Fourier series
\be
u=\sum\limits_{j=0}^{\infty}a_j^{(u)}\cos(j\alpha p_s\tau)+b_j
^{(u)}\sin(j\alpha p_s\tau),\label{14}
\ee
with $p$, $v$, and $z$ being analogous Fourier series with respective
coefficients. Substituting these series in the basic equations (\ref{7}), we get
an infinite number of recurrent relations for the coefficients. A period-1
limit cycle is described by the first terms in the Fourier series, if amplitudes
of higher harmonics can be neglected. The approximated solution at large
detunings can be found (truncating the hierarchy at $j=2$) to be
\begin{equation}
\begin{array}{l}
p\approx p_s-(\delta A_b/2\alpha p_s)\cos2\alpha p_s\tau,\\
u\approx-2\delta A_b\cos\alpha p_s\tau,\\
v\approx -\gamma_a A_b\cos\alpha p_s\tau+2\alpha p_s A_b
\sin\alpha p_s\tau,\label{ballist}\\
z\approx -1+A_b(1+\cos 2\alpha p_s\tau),
\end{array}
\end{equation}
where
\begin{equation}
A_b\equiv \frac {n}{(\delta^2-\alpha^2 p_s^2+n+\gamma_a^2/4)}.  \label{ball}
\end{equation}
The solutions (\ref{ballist}) and (\ref{ball}) are valid under the condition
$n\ll\delta^2-\alpha^2 p_s^2+\gamma_a^2/4$. Numerical simulation shows
that the obtained solutions really work well under the condition of a small
number of photons as compared with the squared detuning. When this condition
breaks down, higher harmonics become significant, and the simple form of
period-1 limit cycles is distorted.

The same procedure has been used to find analytic solutions for small
oscillations of atoms in wells of the optical potential. At $\delta>0$, the
wells are situated at the nodes of the standing wave $\xi=\pi/2+\pi m$
($m=0,1,...$). Assuming a harmonic motion of an atom at the bottom of a well,
$\xi=\pi/2+\pi m+\xi_m\cos\omega_\xi\tau$, we find the approximated solutions for a period-1
limit cycles
\begin{equation}
\begin{array}{lll}
\rho\approx (2\delta A_w\xi_m/\omega_\xi)\sin\omega_\xi\tau,\\
u\approx-2\delta A_w\xi_m\cos\omega_\xi\tau,\\
v\approx -\gamma_a A_w\xi_m\cos\omega_\xi\tau+2\omega_\xi A_w\xi_m
\sin\omega_\xi\tau,\\
z\approx -1+A_w\xi_m^2(1+\cos \omega_\xi\tau),\label{well}
\end{array}
\end{equation}
where the squared frequency of the small oscillations is
\begin{equation}
\label{16}
\begin{displaystyle}
\omega^2_\xi\approx -\frac{\delta^2}{2}-\frac{\gamma_a^2}{8}
+\frac{1}{2}\sqrt{\left(\delta^2+\frac{\gamma^2}{4}\right)^2+8n\alpha|\delta|}
\end{displaystyle}
\end{equation}
and
\begin{equation}
\label{17}
A_w\equiv \frac{n}{\delta^2-\omega^2_\xi+n\xi_m^2+\gamma_a^2/4}.
\end{equation}
The solutions (\ref{well}) -- (\ref{17}) are valid under the following condition:
$n\xi_m^2\ll\delta^2-\omega^2_\xi+\gamma_a^2/4$. Under the limits of its
validicity, the solutions (\ref{well}) -- (\ref{17}) agree well with numerical simulation.
Since the normalized amplitude of small oscillations, $\xi_m$, can be
less than 1, the solutions (\ref{well}) are valid up to the values of the
number of photons of the order of a few thousands.

\subsection{Riddling}

For given values of the recoil frequency $\alpha$ and the spontaneous rate
$\gamma_a$, there are many kinds of attractors when the detuning $\delta$ and
the numbers of photons $n$ are changed. For the same $n$ and $\delta$, different
initial conditions may lead to different attractors, i. e. the system is
sensitively dependent on initial conditions. It should be stressed that it
may happen even under regular synchronous oscillations. We have numerically
found two, three, and more coexisting stable limit cycles with the same and
different periods, whose basins of attraction are riddled by each other. The
basin is said to be riddled \cite{Ott}, if any point in the basin has in its
arbitrary small vicinity points of other attractor basins.

\begin{figure}[htb]
\begin{center}
\epsfxsize=6.5cm
\epsfbox{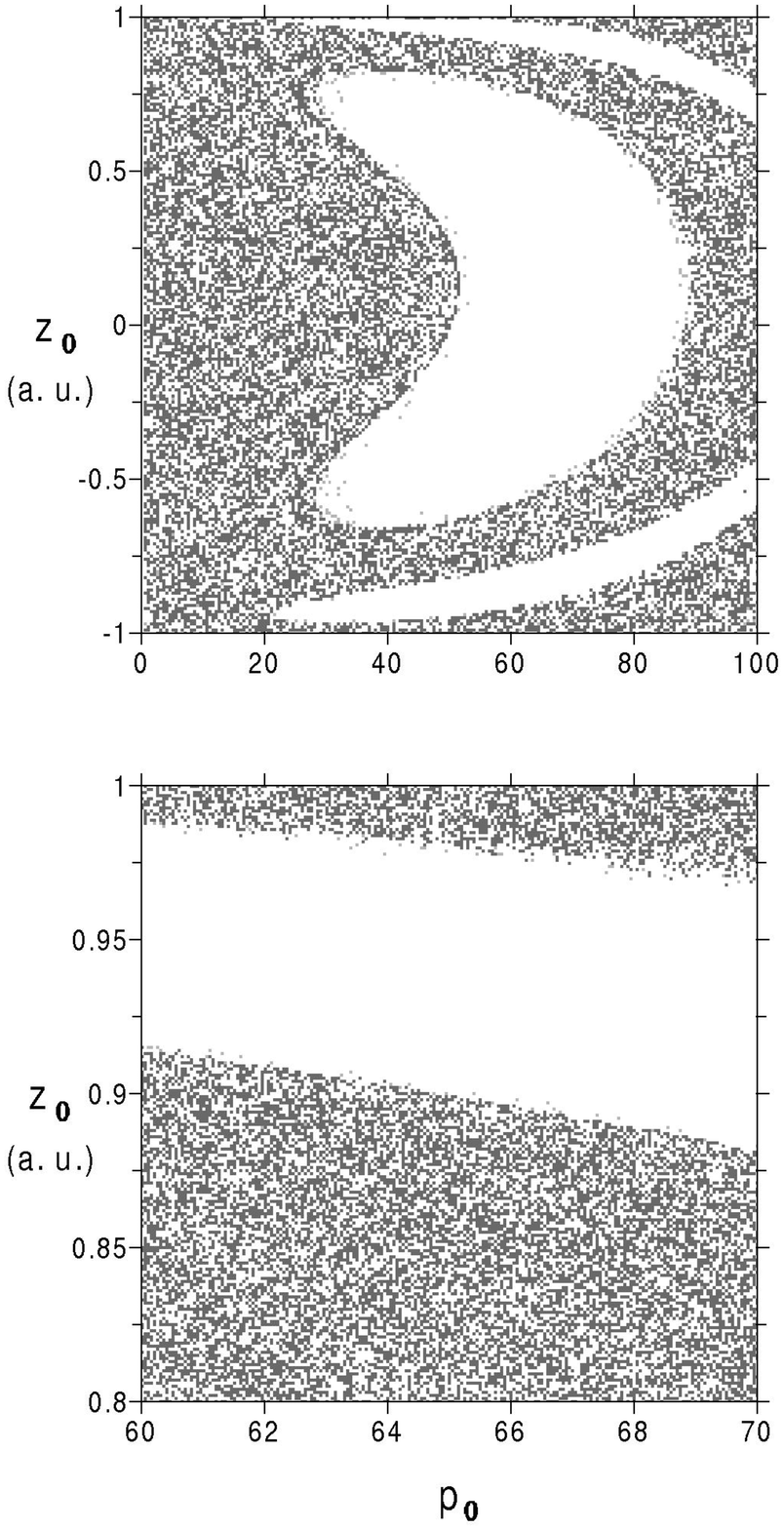}
\end{center}
\caption{\small Riddled basins of attraction in the plane of initial population
inversion $z_0$ and initial momentum $p_0$ (in units $\hbar k_f$):
{\fboxsep=0pt \fbox{\protect\vphantom{A}\textcolor[gray]{1.0}{\vrule width 0.8em depth 0pt}}}~period 1 limit cycle,
{\fboxsep=0pt \fbox{\protect\vphantom{A}\textcolor[gray]{0.75}{\vrule width 0.8em depth 0pt}}}~period 2 limit cycle,
{\fboxsep=0pt \fbox{\protect\vphantom{A}\textcolor[gray]{0.6}{\vrule width 0.8em depth 0pt}}}~period 3 limit cycle.
(a) $200\times200$ grids and (b) $200\times200$ grids giving a magnified view
of a small region in (a).}
\label{fig04}
\end{figure}

In Fig.~4 we illustrate riddling for three coexisting limit cycles of periods 1, 2,
and 3 in the plane of initial atomic momentum $p_0$ and population inversion
$z_0$. Initial conditions on $200\times200$ grid were integrated to find the
respective limit cycles they destinate. Grid points going to the period-3
limit cycles are plotted as dark grey dots, those that go to the period-2
limit cycle --- as light grey dots, and the grid points going to the period-1
limit cycle are left white. As one can see in Fig.~4, grey and white points
are mixed. Since the accuracy of preparing initial conditions is finite, one
cannot predict exactly which limit cycle will be destinated from a given
initial point in the phase space. To confirm the riddling, we magnified a
small region in Fig.~4a ($0.8\leqslant z_0\leqslant1$;
$60\leqslant p_0\leqslant70$) and compute the basins of attraction with the same
grid $200\times200$. The result, shown in Fig.~4b, demonstrates that
riddled basins
may occur not only on chaotic and strange attractors \cite{Ott} but also on
stable limit cycles.

\subsection{Spectra of fluorescence}

The feasible way to detect synchronization in real experiments is to measure
spectra of atomic fluorescence. Let us derive an expression for the atomic
transition dipole moment in the laboratory frame of reference. Assuming the
magnitude $\mu$ of the dipole $d$ to be real, we have $d=\mu\tilde x$,
where $\tilde x$-component of the dipole in the laboratory frame is connected
to the $x$ and $y$ components in the frame, rotating with the frequency $\omega_f$,
by the following way:
\be
\tilde x=x\cos\omega_ft+y\sin\omega_ft.\label{19}
\ee
Our basic equations (\ref{7}) are written in the rotating frame for the dynamical
variables $u\equiv\sqrt{n/2}\ (x+y)$ and $v\equiv\sqrt{n/2}\ (y-x)$. Therefore,
the expression for the atomic dipole moment in the laboratory frame is
\be
\bd
d=\frac{\mu}{\sqrt{2n}}\left[(u-v)\cos\omega_ft-(u+v)\sin\omega_ft\right].\label{20}
\ed
\ee
Using the solutions (\ref{ballist}) for a period-1 limit cycle in the ballistic mode,
it is easily to find that the atomic dipole moment will oscillate with the frequencies
$\omega_f\pm jk_f v_s$, where $ v_s$ is a quasistationary atomic velocity
reached by  atoms after a transition time, and $j=0,1,2,...$. Dipole moments
of ballistic atoms with a period-m limit cycle have the frequencies
$\omega_f\pm jk_f v_s/m\equiv\omega_f\pm j\Omega_0\alpha p_s$. Trapped atoms
oscillating with a period-m
limit cycle, have dipole moments oscillating with the frequencies
$\omega_f\pm j\Omega_\xi/m\equiv\omega_f\pm j\Omega_0\omega_\xi/m$, where
$\omega_\xi$ is given by Eq. (\ref{16}). Fluorescence spectra
both for the atoms, randomly walking in a standing wave and oscillating in wells
in the regime of strange chaotic attractors, should be broadened.

\begin{figure}[htb]
\begin{center}
\epsfxsize=8.5cm
\epsfbox{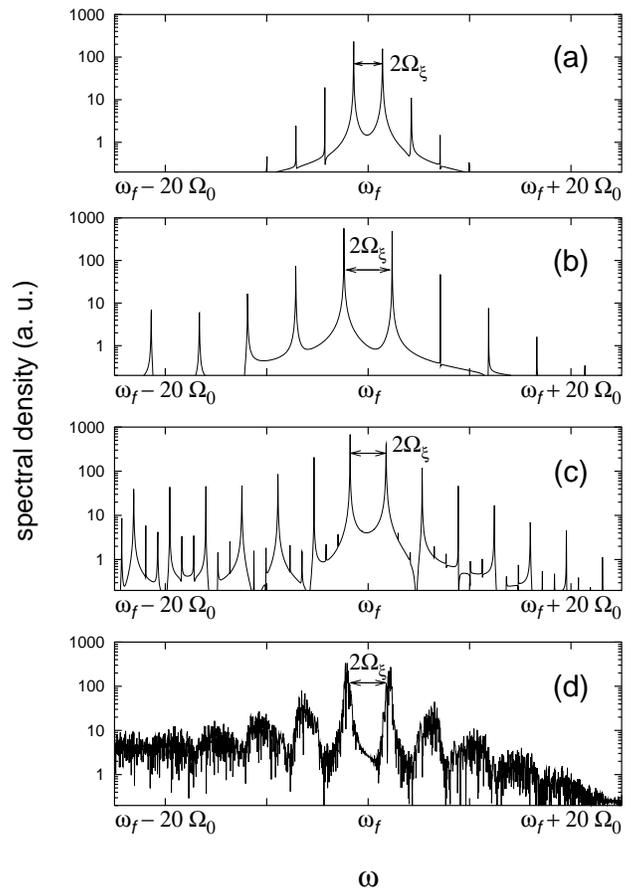}
\end{center}
\caption{\small Fluorescence spectra for trapped atoms in different regimes of
oscillations: at $\delta=24$: (a) period-1 limit cycle with $n=3000$  photons,
(b) period-1 limit cycle with $n=10000$, (c) period-3 limit cycle with
$n=14846$, and (d) chaotic strange attractor with $n=24000$. The frequency
$\omega$ is shown in units of $\Omega_0$.}
\label{fig05}
\end{figure}

In order to check the conclusions mentioned above we integrate numerically
Eq. (\ref{7}) and compute the Fourier spectra of the quantity
$(u-v)\cos\omega_ft-(u+v)\sin\omega_ft$ in different
regimes of oscillations. In Fig.~5 spectra for different limit cycles
for the same parameters as in Fig.~3 are shown. Fig.~5a demonstrate the spectrum
of the quasiharmonic period-1 limit cycle at $n=3000$. In Fig.~5b a more complicated
spectrum of the period-1 limit cycle at $n=10000$ is shown. Fig.~5c demonstrate
the spectrum of the period-3 limit cycle ($n=14846$), and Fig.~5d is a spectrum
of the chaotic strange attractor ($n=24000$). A commom feature of all the spectra
is absence of odd harmonics $\omega_f\pm 2j\Omega_\xi/m $
$(j=0,1,2, ...)$. For quasiharmonic period-1 limit cycles, it can be shown analytically
in the framework of the approach described in \setcounter{abcd}{4}Sec.~
\Roman{abcd}B, that odd harmonics are really absent in the oscillations of
the dipole-moment variables.

All the spectra refer to a
single atom. A multiatom spectrum should be broadened because of a velocity
distribution. However, the effect of velocity grouping, discussed in
\setcounter{abcd}{3}Sec.~
\Roman{abcd}B,
should help to resolve the sidebands of the light emitted by an ensemble of
atoms with limit cycles of low periods.

\section{Bifurcations}
\subsection{Synchronization map}

Our nonlinear dissipative dynamical system demonstrates a rich variety of
dynamical long-term regimes including periodic, quasiperiodic, and chaotic
strange attractors. To illustrate this complexity we plot in Fig.~6
a synchronization map in comparatively small ranges of the detuning and the number
of photons with fixed values of the other control parameters, $\alpha=0.01$,
$\gamma_a=0.3$, and the same initial conditions. White color corresponds
to period-1 limit cycles, nuances of grey color --- to limit cycles of higher
periods, and black color --- to deterministic chaos. As the number of photons
increases, higher harmonics in the spectrum become more and more prominent, and
the approximated solutions (\ref{ballist}) -- (\ref{17}) for period-1 limit cycles are no more valid. It is
seen from the map that the limit cycles with higher periods appear practically
for random values of the control parameters but inside prominent parallel bands
separated by zones with period-1 limit cycles, exclusively. The border of chaos
in Fig.~6 is jagged, and there exist nearby small ``islands'' of
chaotic motion submerged into the ``sea'' of period-1 limit cycle motion.

\begin{figure}[htb]
\begin{center}
\epsfxsize=8cm
\epsfbox{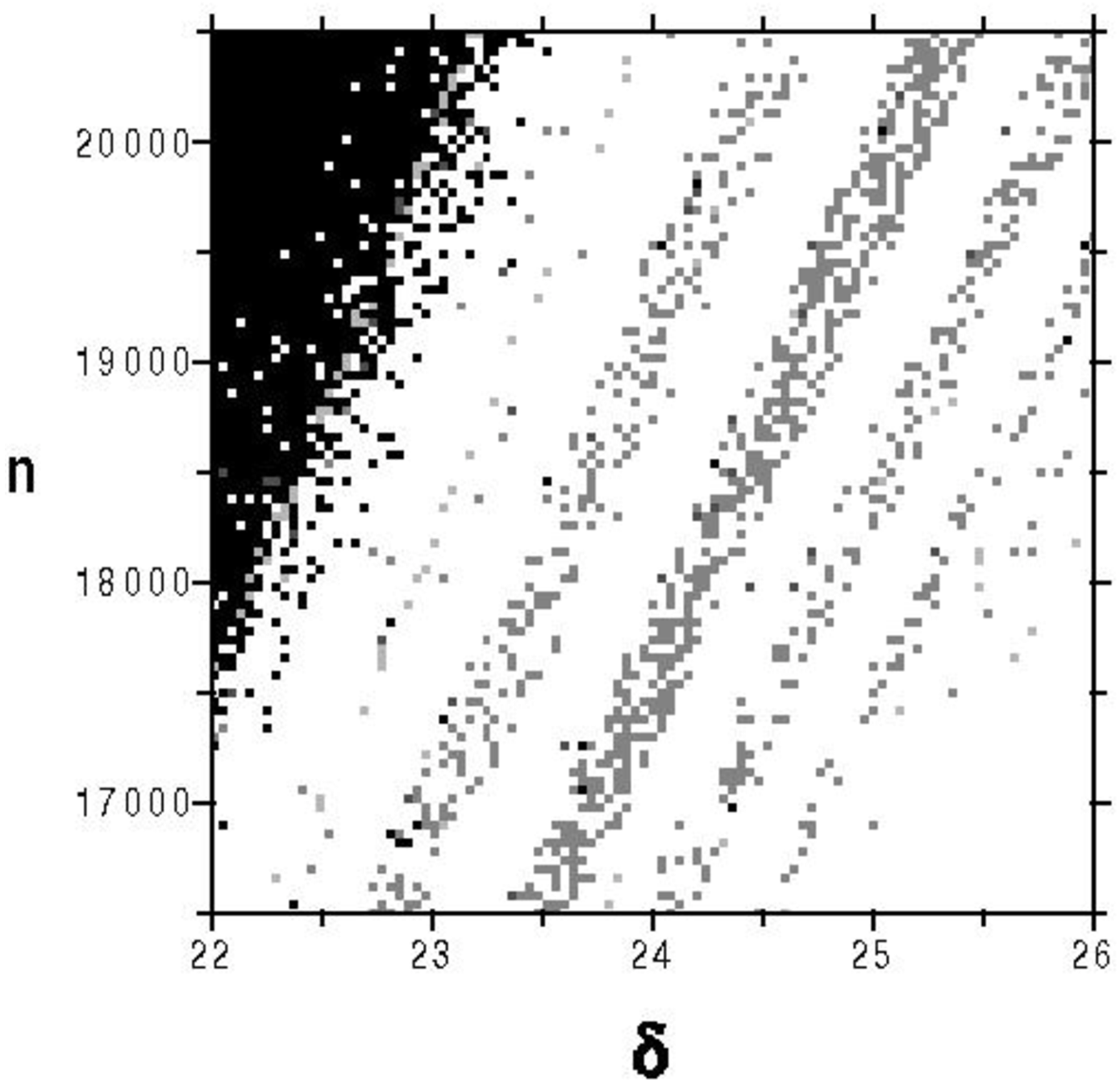}
\end{center}
\caption{\small Synchronization map in the plane of the average numbers of photons
$n$ and the detuning $\delta$ (in units of $\Omega_0$):
{\fboxsep=0pt \fbox{\protect\vphantom{A}\textcolor[gray]{1.0}{\vrule width 0.8em depth 0pt}}}~period 1 limit cycle,
{\fboxsep=0pt \fbox{\protect\vphantom{A}\textcolor[gray]{0.75}{\vrule width 0.8em depth 0pt}}}~period 2 limit cycle,
{\fboxsep=0pt \fbox{\protect\vphantom{A}\textcolor[gray]{0.6}{\vrule width 0.8em depth 0pt}}}~period 3 limit cycle,
{\fboxsep=0pt \fbox{\protect\vphantom{A}\textcolor[gray]{0.4}{\vrule width 0.8em depth 0pt}}}~limit cycles with
periods from 4 to 12, and
{\fboxsep=0pt \fbox{\protect\vphantom{A}\textcolor[gray]{0.0}{\vrule width 0.8em depth 0pt}}}~chaos.}
\label{fig06}
\end{figure}

\begin{figure}[htb]
\begin{center}
\epsfxsize=8.5cm
\epsfbox{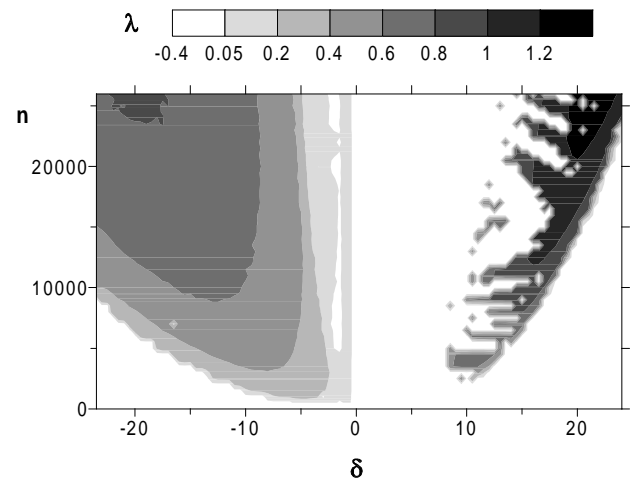}
\end{center}
\caption{\small Dependence of the maximal Lyapunov exponent $\lambda$ on the average
number of photons $n$ and the detuning $\delta$ ($\lambda$ and $\delta$
are shown in units of $\Omega_0$).}
\label{fig07}
\end{figure}

The case of chaos we identify computing the maximal Lyapunov exponent $\lambda$
which characterizes a mean velocity of diverging of initially closed trajectories
in the phase space. The dependence of $\lambda$ on the parameters $\delta$ and
$n$ is shown in Fig.~7 where color modulates values of $\lambda$.
If $\lambda>0$ the atomic oscillations are chaotic in a sense of sensitive
dependence on initial conditions. A prominent asymmetry with respect to sign
of the detuning $\delta$ is explained by different character of the force acting
on atoms when $\delta>0$ and $\delta<0$
(see \setcounter{abcd}{3}Sec.~\Roman{abcd}B).

\subsection{Bifurcation diagram}

A general approach in studying the complexity of nonlinear dynamical systems
involves investigating qualitative changes in their dynamics as system's
control parameters are varied. Sudden changes in the dynamics, known as
bifurcations, occur when one of the control parameter crosses a critical
value. The best way to visualize it is to compute so-called bifurcation diagrams.
The bifurcation diagram in Fig.~8 shows dependence of negative values of the
$v$-component of the atomic dipole of trapped atoms at the moments of
time when $u=0$ on the
average number of photons $n$ at fixed initial conditions and $\delta=24$.

\begin{figure}[htb]
\begin{center}
\epsfxsize=8cm
\epsfbox{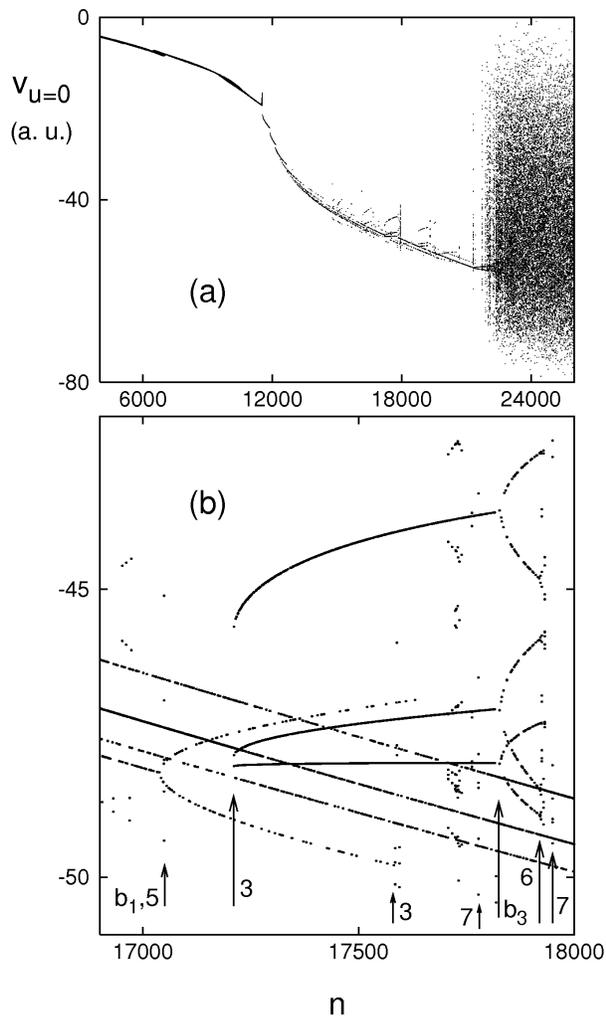}
\end{center}
\caption{\small Attractor bifurcation diagrams at $\delta=24$: (a) a general view and
(b) a cascade of bifurcations. Arrows $b_{1, 3}$ indicate doubling of period-1
and period-3 limit cycles, respectively; 3, 5, and 7 --- period-3, 5, and 7
limit cycles, respectively; 6 --- period doubling bifurcation $3 \to 6$.}
\label{fig08}
\end{figure}

A general view in Fig.~8a demonstrates a complicated scenario of
bifurcations on the road to chaos. In the range $n\lesssim12000$, there exists
a single stable limit cycle with period 1. At $12000 \lesssim n\lesssim 14000$, this
cycle begins to multiply, i.e. there coexist in the phase space a few different
period-1 limit cycles. Starting with a given initial state, an atom after a
transition time reaches, of course, a definite period-1 limit cycle, but small
changes in initial state may lead to another period-1 limit cycle. In other words,
we know that in this range of $n$ a period-1 limit cycle is settled after
a transition time, but one cannot predict practically which one is reached.
In the range $14000 \lesssim n\lesssim 21000$, a few cascades of bifurcations
occur where limit cycles of higher periods appear.

A more detailed view in Fig.~8b helps to see a fine structure of one
of the bifurcation cascades. We see that at $n\lesssim17050$ practically all
the points belong to four parallel lines. At a given value of $n$, only one value
of $v$ exists, but under small changes in $n$ values of $v$
jump randomly between the four lines. There really coexist in the phase
space four different limit cycles of period 1 with their own basins of attraction.
A pitchfork $b_1$ at $n\simeq17050$ means not a period doubling bifurcation
but birth of two new limit cycles with period 1 from the old one.
With $n$ increasing further, real bifurcations with period $m$ occur. We
indicate them in Fig.~8b by the respective numbers. A prominent
bifurcation of period 3 occurs at $n\simeq 17200$ when a period-3 limit cycle
appears suddenly. Cycles of period 5 and 7 are also identified. Famous period
doubling bifurcation are rather rare. One of them, corresponding to appearance
a period-6 instead of a period-3 limit cycle, is indicated in Fig.~8b
as ``6''. In all the range, $16900\lesssim n\lesssim 18000$, there coexist, at least, three
period-1 limit cycles.

\begin{figure}[htb]
\begin{center}
\epsfxsize=8cm
\epsfbox{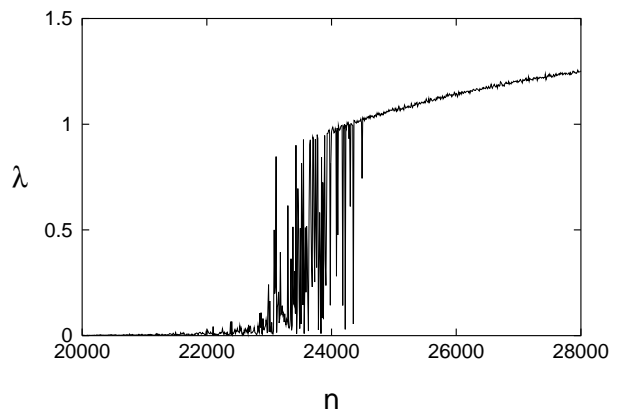}
\end{center}
\caption{\small Maximal Lyapunov exponent $\lambda$ (in units of $\Omega_0$) as a
function of the average number of photons $n$ at $\delta=24$.}
\label{fig09}
\end{figure}

\begin{figure}[htb]
\begin{center}
\epsfxsize=8.5cm
\epsfbox{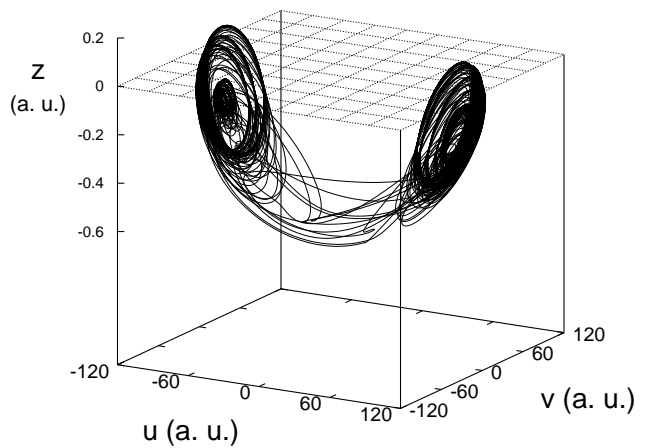}
\end{center}
\caption{\small Strange chaotic attractor in the Bloch sphere at $\delta=24$ and
$n=24000$.}
\label{fig10}
\end{figure}

It is seen from Fig.~8a that, beginning with $n\approx22000$, a
complicated intermittent motion takes place. To confirm it, we compute in detail
the dependence of the maximal Lyapunov exponent $\lambda$ on the average
number of photons $n$ at $\delta=24$. The plot in Fig.~9
demonstrates an intermittent road to a ``stable'' chaotic mode that is nothing
more than a strange chaotic attractor. Its view in the Bloch space is shown
in Fig.~10 with the parameter values $\delta=24$ and $n=24000$.
Projections of this attractor on the planes ($v, u$) and ($\xi, z$) are shown
in Fig.~3c. Its chaoticity is confirmed by computing tha maximal the
Lyapunov exponent to be $\lambda\approx 1$, and its strangeness --- by computing
the Hausdorff dimension $d_F\approx2.7$.

\subsection{Influence of noise}

The cascade of bifurcations in Fig.~8b is very complicated and its
detailed analysis in the whole parameter space is practically impossible. We
cannot expect, of course, that all the fine details of this cascade are relevant
for real atoms in a standing wave. In order to know which limit cycles and
bifurcations are expected to survive under an inevitable noise in real experiments,
we incorporate a stochastic force into the right side of the second equation
in the set (\ref{7}) and compute again the bifurcation diagram shown in
Fig.~8b at different levels of noise. Noise destroys the fine structure
but the prominent period-1 and period-3 limit cycles with their characteristic
properties do not disappear. In Fig.~11 we demonstrate the bifurcation
diagram in the range $16900\lesssim n\lesssim 18000$ under a weak broadband
noise. Estimating the frequency of nonlinear oscillations of trapped atoms
to be in the range $\omega_\xi\simeq 0.3 - 2$, we  model the broadband
noise by a linear combination of 100 harmonic functions with
random phases and equidistantly distributed frequencies in the range from
0.05 to 5.

\begin{figure}[htb]
\begin{center}
\epsfxsize=8cm
\epsfbox{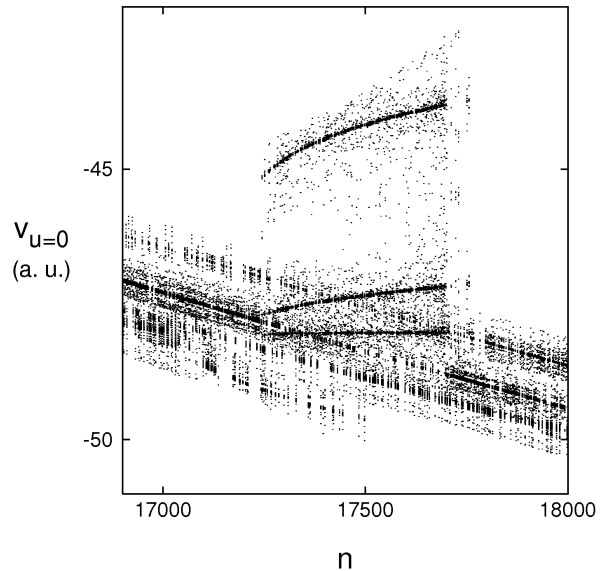}
\end{center}
\caption{\small Attractor bifurcation diagram at a weak noise. The other parameters
and initial conditions are the same as in Fig.~8.}
\label{fig11}
\end{figure}

\section{Fractal random walking of atoms}

In \setcounter{abcd}{3}Sec.~\Roman{abcd} we have considered mechanical effects
in regular atomic motion: acceleration, deceleration, velocity grouping,
and trapping of atoms, all of which were characterized by negative or zero
values of the maximal Lyapunov exponent $\lambda$. The Lyapunov map in Fig.~7
demonstrates that there exist wide ranges of the control parameters, $n$ and
$\delta$, where atomic motion is expected to be exponentially sensitive
to small changes in initial conditions. In the chaotic regime there are also
exist oscillations in wells of the optical potential and ballistic flights
with an average velocity $ p_s$. However, instead of stable limit cycles,
there appear chaotic attractors (see Figs.~3c and 10), and the instant
atomic velocity in the ballistic mode oscillates around $ p_s$ in
an irregular way. In Fig.~12a we plot an example of a chaotized velocity
\begin{figure}[htb]
\begin{center}
\epsfxsize=8.5cm
\epsfbox{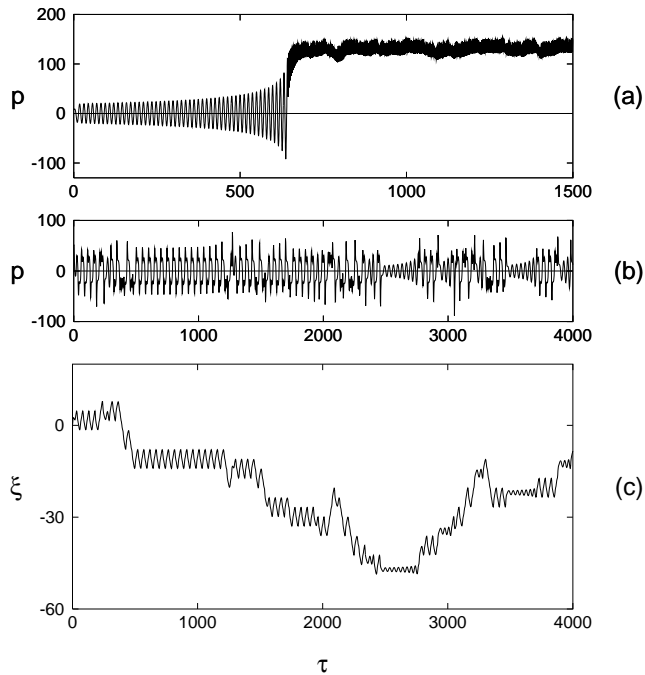}
\end{center}
\caption{\small Chaotic atomic motion: (a) chaotic velocity grouping at $\delta=-24$,
(b) and (c) random walking at $\delta=-3$. Atomic position in units $k_f^{-1}$ and
momentum $p$ in units $\hbar k_f$ as functions of time. Here $n=24000$.}
\label{fig12}
\end{figure}
grouping (at $\delta =-24$, $n=24000$, and $p_0=8$) that should be compared
with the velocity grouping in regular regime in Fig.~2a. In difference from
the regular velocity grouping, the momentum $ p_s$, averaged over
time of flying between two successive nodes of the standing wave,
is not constant but oscillates slightly in an irregular way. With sufficiently
large $ p_s$ it does not matter, in principle, but in some ranges of
the control parameters a cardinally new type of motion takes place, when
atomic motion is neither ballistic nor oscillating in wells but resembles
strongly a random walking. In this regime an initially trapped atom
oscillates in a well for a while, then it leaves the well for a ballistic
flight and can be trapped in another well, etc. Random walking appears when there
exist such zeroes of the friction force $F$ where the grouping
momentum $ p_s$ is close to the critical momentum $p_{cr}$ (see
\setcounter{abcd}{3}Sec.~\Roman{abcd}B). In the chaotic regime ($\lambda >0$),
$ p_s$  varies in an irregular way being either greater
or smaller than $p_{cr}$. As an example, we demonstrate in Fig.~12b and c
a random walking (at $\delta =-3$, $n=24000$, and $p_0 = 50$) with a rather
small grouping momentum $ p_s \approx 30$.

\begin{figure}[htb]
\begin{center}
\epsfxsize=7.5cm
\epsfbox{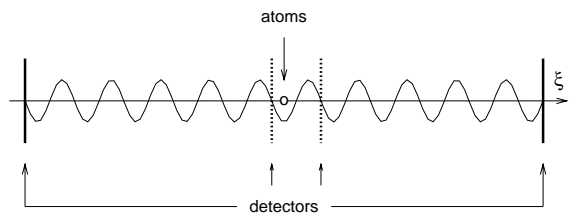}
\end{center}
\caption{\small Schematic diagram showing scattering of atoms at the standing wave
when the space between detectors is equal to two (dashed lines) and twenty
one wavelengths (bold lines).}
\label{fig11}
\end{figure}

Interesting properties of randomly walking atoms will be demonstrated with
the following numerical experiment, a scheme of which is shown in
Fig.~13. Atoms, one by one, are placed at the point $\xi=0$ with the
same initial momentum $p_0=50$ along the cavity axis. We compute the time
$T$ the atoms need to reach one of the detectors varying one of the control
parameters, the average number of photons, $n$, or the detuning $\delta$, and with
the other parameters and initial conditions to be the same. To reduce
numerical efforts, we restrict ourselves by 1D-geometry.
The analogous numerical experiments for Hamiltonian atomic motion in a high-Q
cavity have been done in Ref. \cite{AP03} for a classical radiation field
and in Refs. \cite{PU03, P04} for a quantized field in the Fock and coherent
states of light. Self-similar hierarchical structures in the dependence $T(p_0)$
have been found in these papers. It seems to be more practical to compute the
exit-time function in dependence on one of the parameters ($n$ or $\delta$)
that is more easy to control than the initial atomic momentum.

\begin{figure}[htb]
\begin{center}
\epsfxsize=8.5cm
\epsfbox{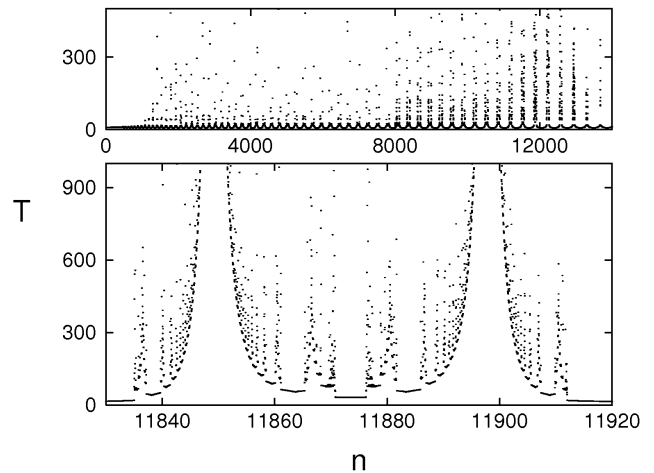}
\end{center}
\caption{\small Dependence of the exit time $T$ (in units $\Omega_0^{-1}$) on the
number of photons $n$ with different resolutions at $\delta=-3$ and $\gamma_a=0.3$.}
\label{fig14}
\end{figure}

The exit-time function $T(n)$ has been computed at $\delta =-3$
with two detectors placed at the distance of two lengths of the standing wave
(dashed lines in Fig.~13 indicate places of the detectors in this case).
Fig.~14 demonstrates an intermittency of smooth intervals and complicated
structures that cannot be resolved in principle no matter how large the
magnification factor. The lower fragment of Fig.~14 is a zoom of the
respective interval of the upper fragment.
We see that even such a small zone of the atom-field interaction is
enough to generate a complicated self-similar structure.

\begin{figure*}[p]
\begin{center}
\epsfxsize=17cm
\epsfbox{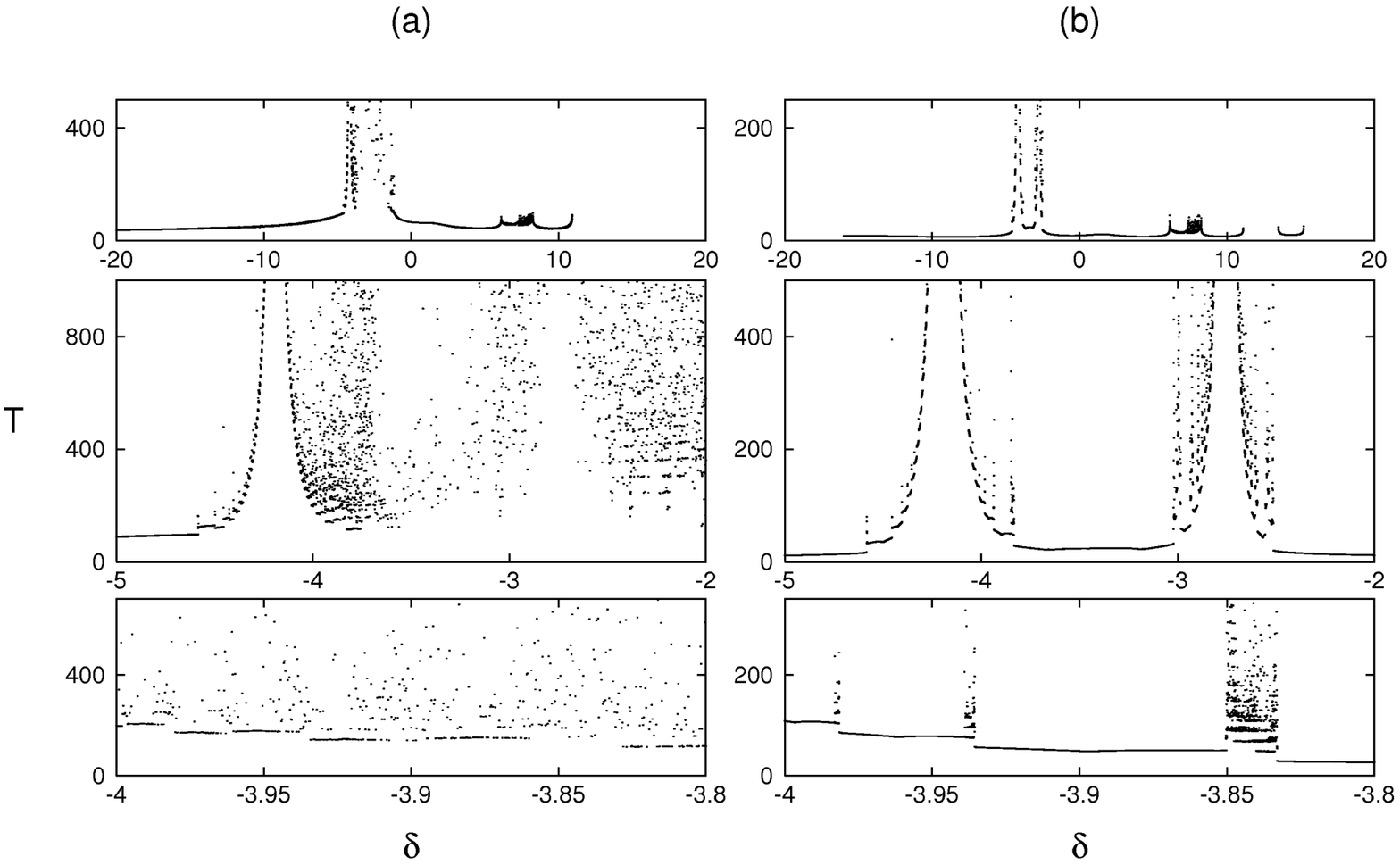}
\end{center}
\caption{\small Dependence of the exit time $T$ (in units $\Omega_0^{-1}$) on the
detuning $\delta$ (in units of $\Omega_0$) with different resolutions at
$n=11880$ and $\gamma_a=0.3$: (a) the space between detectors is 21
wavelengths and (b) 2 wavelengths.}
\label{fig15}
\end{figure*}

In order to prove that fractality is a robust property of randomly walking
atoms, we compute the exit times $T$ in dependence on the detuning
$\delta$ with the space between detectors equal to 21 wavelengths.
The result is shown in Fig.~15a at $n=11880$ photons. Comparing it with
the same function $T(\delta)$, but computed with two wavelengths,
we see that the structure in Fig.~15b is more ordered demonstrating
under zooming an intermittency of smooth and irregular parts with
singularities.

The exit time $T$, corresponding
to both the smooth and unresolved intervals, increases in average with increasing
the magnification factor. There are also exist atoms never reaching the
detectors in spite of the fact that they have no obvious energy restrictions
to leave the small space between two detectors. There is an infinite number of
unstable periodic orbits and chaotic orbits. The measure of such orbits is,
of course, zero, and they cannot be exactly detected even in numerical
experiments. However, such a sensitive dependence of the measured characteristic
on the detuning should manifest itself in wild oscillations of the exit
times in an arbitrary small neighbourhood of such values of $\delta$ where the
scattering function $T(\delta)$ is singular, and the range of variation of $T$
should not decrease to zero as the size of such a singular neighbourhood is
reduced. The scattering function is singular on a Cantor set of values
of the control parameter. It means that a relatively small uncertainty in the
control parameters (or in initial values of dynamical variables) can often
make prediction of the exit time practically impossible.
The fractal structure has been observed only with randomly walking atoms,
i.e. at negative detunings. We have found complicated structures in the
scattering function at $\delta >0$, however under a magnification they
appear to be smooth and without singularities.

\begin{figure*}[p]
\begin{center}
\epsfxsize=17cm
\epsfbox{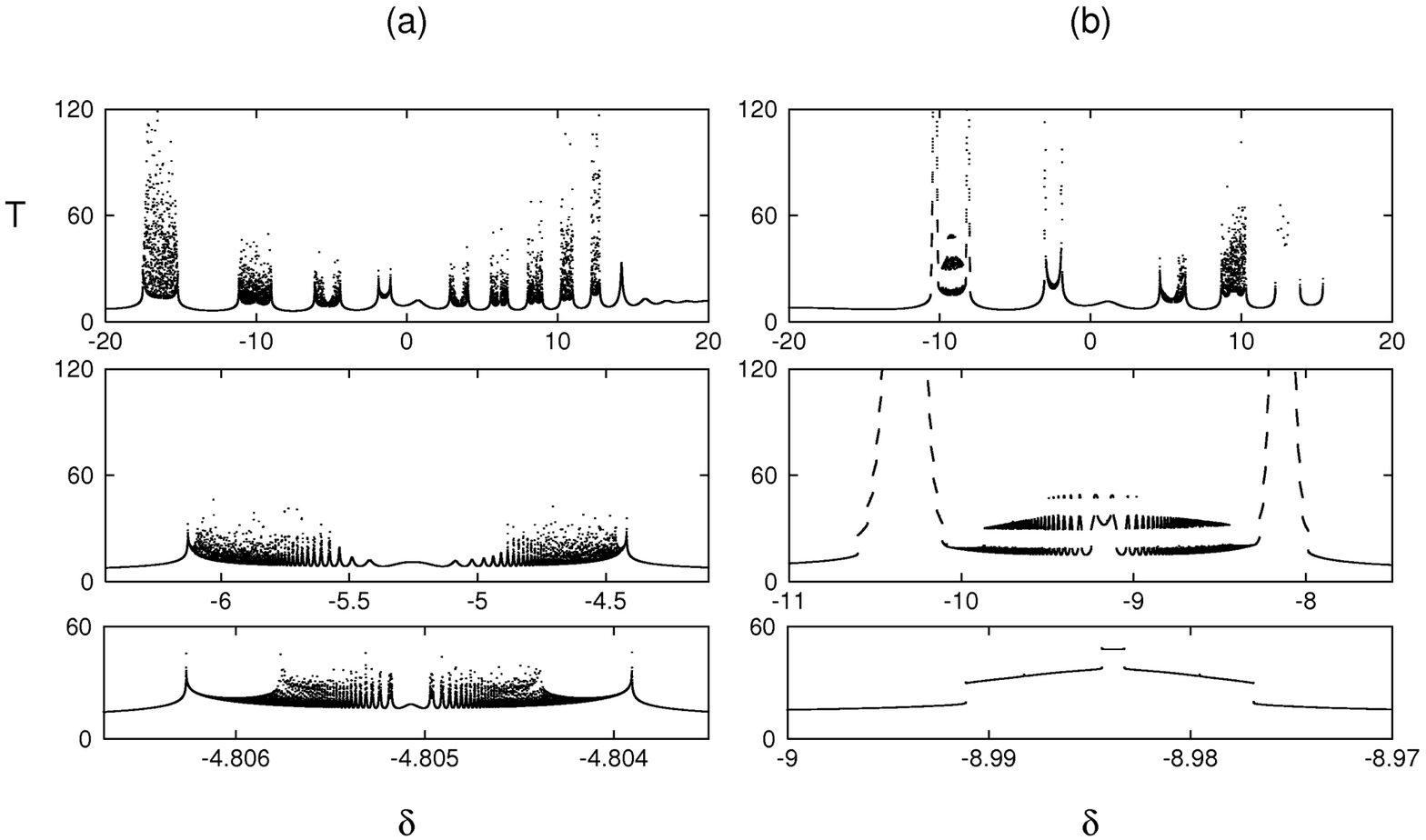}
\end{center}
\caption{\small Comparing Hamiltonian fractal with $\gamma_a=0$ (a) and dissipative
fractal with $\gamma_a=0.2$ (b).}
\label{fig16}
\end{figure*}

To clarify the role of dissipation in forming fractal structures in the numerical
scattering experiments, we plot in Fig.~16a the same dependence
$T(\delta)$ but without dissipation, i.e. $\gamma_a=0$. The fractal in
Fig.~16a is a Hamiltonian one, and it possesses a beautiful self-similar
structure with intermittent smooth and singular zones. Comparing
Figs.~16a with $\gamma_a =0$ and 15b with $\gamma_a =0.3$, we see that
the main part of this
structure disappears under influence of dissipation. Instead of singularities,
one can see smooth intervals in Fig.~15b. In order to demonstrate
how in occurs, we plot in Fig.~16b the function $T(\delta)$ under
the same conditions but with a reduced dissipation, $\gamma_a=0.2$. In places
of unresolved structures in Fig.~16a, we see some structures that, however,
are resolved under a magnification into smooth intervals. We may conclude that
spontaneous relaxation changes the fractal structure of the Hamiltonian scattering
function. It smoothes second-order self-similar structures of the Hamiltonian
fractal \cite{AP03}, but their border singular points become fractal-like
dissipative structures.

\section{Conclusion}

We have performed a theoretical and numerical study of the phenomenon of
synchronization between internal and external degrees of freedom of atoms
in a standing laser wave and of related topics. The character of the
center-of-mass atomic motion can be understood with the help of the
dependence of the friction force on the average momentum
$F(|\bar p|)$. Zeroes of this function are either attractors or
repellors. As a result, atoms tend to be grouped around respective
values in the velocity space, the effect known as velocity grouping.

Synchronization is a common rhythm of co-existence of mechanical
oscillations of atomic center-of-mass motion and internal Rabi
oscillations that manifests itself in the phase space as limit cycles
with different periods and riddled basins of attraction. We have derived
a set of coupled equations for external and internal  atomic variables
and obtained approximated analytical solutions for period-1 limit cycles
as for atoms, trapped in wells of the optical potential, as for
ballistic atoms. The feasible way to detect  synchronization in real
experiments is to measure spectra of atomic fluorescence which for
a period-m limit cycle should contain sideband frequencies,
$\omega_f\pm jk_f v_s/m$ (ballistic atoms), or
$\omega_f\pm j\Omega_\xi /m$ (trapped atoms). Due to the velocity grouping,
limit cycles can be detected in spectra of light emitted by an ensemble
of atoms with an initial velocity distribution.

Increasing the number of photons in the standing wave $n$, we have found
 cascades of bifurcations with higher periods to appear which eventually
settle in a ``stable'' chaotic strange attractor with a positive value
of the maximal Lyapunov exponent $\lambda$ and a fractional  Hausdorff
dimension. A broadband noise, added to the equations of motion, destroys
a fine structure of the cascades, but the prominent period-1 and period-3
limit cycles survive under a weak noise.

Computing the Lyapunov map, we have found wide ranges of the control
parameters, $n$ and the detuning $\delta$, where $\lambda >0$, and the
atomic motion is expected to be chaotic either in the form of strange
attractors in the wells or as flights with irregular oscillations of the
velocity. Moreover, there exists a regime where the atomic motion is
neither one of them but resembles strongly a random deterministic walking
with erratically jumping atoms. The random walking is fractal in the sense
that  a scattering function is very sensitive to small changes in the
control parameters or initial conditions and has a self-similar structure
with singularities on a Cantor set of values of one of the control parameters.
We have demonstrated it by computing a time of exit of atoms from a given
space of the standing wave in dependence on $n$ or $\delta$.

\section{Acknowledgments}

The work was financially supported by the Program of the Prezidium of
the Russian Academy of Sciences, by the Russian Foundation for
Basic Research (project 02-02-17796) and by the Far-Eastern Division
of the Russian Academy of Sciences.

\end{document}